\documentclass[10 pt,fleqn]{ieeeconf}
\IEEEoverridecommandlockouts                              
\usepackage{cite}
\usepackage{amsmath,amssymb,amsfonts}
\usepackage{algorithmic}
\usepackage{graphicx}
\usepackage{textcomp}

\usepackage{algorithm}
\usepackage{algorithmic}

\DeclareMathOperator{\rank}{rank}

\DeclareMathOperator{\range}{range}
\DeclareMathOperator{\cov}{cov}
\DeclareMathOperator*{\minimise}{minimise}
\DeclareMathOperator{\st}{subject\ to:}     

\newtheorem{theorem}{Theorem}
\newtheorem{lemma}[theorem]{Lemma}
\newtheorem{problem}[theorem]{Problem}

\newcommand{\Uset}{{\cal U}}
\newcommand{\Yset}{{\cal Y}}
\newcommand{\R}{{\cal R}}

\newcommand{\Splusn}{{\cal S}_{+}^{\,n}}
\newcommand{\Splus}{{\cal S}_{+}}
\newcommand{\Sn}{{\cal S}^{\,n}}

\newcommand{\Exp}[1]{{\cal E}\!\left\{#1\right\}}  
\newcommand{\var}[1]{\operatorname{var}\!\left(#1\right)}

\newcommand{\ycost}{P}
\newcommand{\ucost}{R}

\newcommand{\Tf}{T_{\text{f}}}
\newcommand{\Tp}{T_{\text{p}}}

\newcommand{\nin}{n_u}
\newcommand{\ny}{n_y}
\newcommand{\nx}{n_x}

\newcommand{\ud}{u^{\text{d}}} 
\newcommand{\yd}{y^{\text{d}}} 

\newcommand{\up}{u_{\text{p}}} 
\newcommand{\yp}{y_{\text{p}}}
\newcommand{\ypmeas}{y_{\text{p,meas}}}
\newcommand{\ymeas}{y_{\text{meas}}}

\newcommand{\uf}{u_{\text{f}}}
\newcommand{\yf}{y_{\text{f}}}

\newcommand{\yfest}{\hat{y}_{\text{f}}}

\newcommand{\xnp}{x_{\text{np}}}
\newcommand{\xyest}{\hat{x}_{y}}
\newcommand{\xu}{x_u}
\newcommand{\xy}{x_y}

\newcommand{\gp}{g_{\text{p}}}

\newcommand{\gf}{g_{\text{f}}}

\newcommand{\Hu}{H_u}
\newcommand{\Hy}{H_y}

\newcommand{\Hup}{H_{u\text{p}}}
\newcommand{\Hyp}{H_{y\text{p}}}
\newcommand{\Huf}{H_{u\text{f}}}
\newcommand{\Hyf}{H_{y\text{f}}}

\newcommand{\Hyup}{H_{yu\text{p}}}
\newcommand{\Hyuf}{H_{yu\text{f}}}

\newcommand{\Lp}{L_{\text{p}}}
\newcommand{\Lf}{L_{\text{f}}}
\newcommand{\Lup}{L_{u\text{p}}}
\newcommand{\Lyp}{L_{y\text{p}}}

\newcommand{\Lyup}{L_{yu\text{p}}}
\newcommand{\Lyf}{L_{y\text{f}}}
\newcommand{\Luf}{L_{u\text{f}}}
\newcommand{\Lyuf}{L_{yu\text{f}}}

\newcommand{\Qp}{Q_{\text{p}}}
\newcommand{\Qf}{Q_{\text{f}}}
\newcommand{\Qup}{Q_{u\text{p}}}
\newcommand{\Qyp}{Q_{y\text{p}}}

\newcommand{\Qyuf}{Q_{yu\text{f}}}
\newcommand{\Qnp}{Q_{\text{np}}}
\newcommand{\Qnf}{Q_{\text{nf}}}

\newcommand{\Inu}{I_{n_u}}
\newcommand{\Inx}{I_{n_x}}

\newcommand{\ITp}{I_{\Tp}}

\newcommand{\Suu}{S_{uu}}
\newcommand{\Suy}{S_{uy}}
\newcommand{\Syu}{S_{yu}}
\newcommand{\Syy}{S_{yy}}

\newcommand{\Eup}{E_{u\text{p}}}
\newcommand{\Eyup}{E_{yu\text{p}}}
\newcommand{\Eyp}{E_{y\text{p}}}
\newcommand{\Euf}{E_{u\text{f}}}

\newcommand{\Exy}{E_{x_y}}

\newcommand{\SigmaV}{\Sigma_V}

\newcommand{\Psixy}{\Psi}

\newcommand{\zetaf}{\zeta_{\text{f}}}
\newcommand{\zetafest}{\hat{\zeta}_{\text{f}}}
\newcommand{\nup}{\nu_{\text{p}}}
\newcommand{\nupmeas}{\nu_{\text{p,meas}}}

\makeatletter
\def\qed{%
   {%
      \unskip
      \nobreak \hfil
      \penalty 50               
      \hskip 3em                
      \null \nobreak \hfil
      $\Box$
      \parfillskip=\z@skip
      \finalhyphendemerits=\z@  
      \endgraf                  
   }}
\makeatother

\title{\LARGE \bf
Optimal Data-Driven Prediction and Predictive Control using Signal Matrix Models
}
\author{Roy S.\ Smith, Mohamed Abdalmoaty, \& Mingzhou Yin
\thanks{This work was supported by the NCCR Automation Grant 51NF40\_180545 
funded by the Swiss National Science Foundation.}
\thanks{The authors are with the Automatic Control Laboratory, Swiss Federal Institute of Technology (ETH Zurich),
Physikstrasse 3, 8092 Z\"{u}rich, Switzerland, {\tt\small \{rsmith,mabdalmoaty,myin\}} {\tt\small @control.ee.ethz.ch}.}%
}

\begin{document}

\maketitle
\thispagestyle{empty}
\pagestyle{empty}


\begin{abstract}
Data-driven control uses a past signal trajectory to characterise the input-output
behaviour of a system.   Willems' lemma provides a data-based prediction model
allowing a control designer to bypass the step of identifying a state-space or transfer
function model.   This paper provides a more parsimonious formulation of Willems'
lemma that separates the model into initial condition matching and predictive control
design parts.  This avoids the need for regularisers in the predictive control problem that
are found in other data-driven predictive control methods.   It also gives 
a closed form expression for the optimal (minimum variance) unbiased predictor of
the future output trajectory and applies it for predictive control.   Simulation comparisons
illustrate very good control performance.
\end{abstract}


\section{Introduction}

Model-based control design has a long history.   Obtaining a suitable model can be a challenging task and 
is often addressed by a combination of first principles modeling and identification from data.   The modeling and
design steps are not independent and the best model depends strongly on the control design objectives.
The recent resurgence in learning methods has driven a renewed interest in designing control systems 
directly from the measured system signal data.   One potential benefit of these data-driven methods
is that bypassing the modeling step in the design process might allow the process to be simplified, or 
even semi-automated, avoiding the costly and time-consuming identification step.   However many
of the experiment design considerations that go into system identification process arise in obtaining
signal data for the data-driven methods.    

This class of data-driven control should not be considered as being ``model-free''.   Rather, the
parameterisation of the model is in terms of a set of past data.   This data-based model is used
for prediction of the system responses to future control input signals.
Our term for this class of models is Signal Matrix Models (SMM).

We start from the framework used in the current data-driven
control work (summarised in \cite{verheijen:2023a,berberich:2021a}), and formulate
the problem as decoupled prediction and predictive control steps.  This allows us to pose and solve
optimal prediction problems which are tailored to the predictive control objective.   One major advantage is 
that regularisation is not required for the prediction and control steps, in contrast to the current
data-driven approaches. 

Our development will bring out the strong connection between Subspace Predictive Control 
(SPC)~\cite{favoreel:1999a} and the more recent data-driven methods based on Willems'
Lemma~\cite{willems:2005c,fiedler:2021a}).    Data Enabled Predictive Control 
(DeePC)~\cite{coulson:2018a} is a prototypical approach and uses a single vector variable
to parametrise both the initial condition matching, or equivalently the prediction, and the 
predictive control steps.  Achieving the best trade-off between these conflicting conditions is
achieved via regularisation.   

The method developed by \cite{breschi:2023a, breschi:2023b}, known as $\gamma$-DDPC, 
reformulates the characterisation of the dynamics and provides a different regularisation 
approach.  One choice of regularisers gives an equivalence with DeePC.   Generalised
Data-Driven Predictive Control (GDPC), described in \cite{lazar:2023a},  splits the control
input in two with one part using the SPC prediction methods to predict the output response
and the other optimised online to satisfy the control objectives.
GDPC also requires regularisation to achieve the prediction-control 
trade-off for the optimised part of the control input signal.

High fidelity simulations, or digital twins, generate ideal data for this approach.  If 
dedicated experiments are possible noise and excitation problems can be addressed
through experiment design.  Data taken from historical operation measurements can
have both noise and persistency of excitation problems.    The regularisation approaches
of DeePC, $\gamma$-DDPC, and GDPC, can help ameliorate the noise problem, but 
a theoretical basis for this is incomplete.  The work in~\cite{Yin:2023a}
develops maximum likelihood methods for the case where the SMM is corrupted by
stochastic noise.  Its use in predictive control requires some heuristics.

Our approach is somewhat different.  We will consider $\Hy$
to be noise-free and derive the optimal predictor to be used in the predictive control step.
In the case that there is noise in $\Hy$ this amounts to considering a high-order
model---one that fits the noise in addition to the system response.   The method given 
here can be further improved by applying denoising methods to the SMM
matrices (see \cite{yin:2021a}).

\subsection{Notation}
\label{ss:notation}

The set of (strictly) positive definite matrices, of dimension $n$ is denoted by ($\Splusn$) $\Sn$. 
We denote the weighted $l_2$-norm of a vector $x$ by $\|x\|_P = (x^T P x)^{1/2}$
where $P \in \Splusn$.   
 A random vector $x$,  has expectation $\Exp{x}$ and covariance $\cov(x)$.
The range space of a matrix $H$ is $\range(H)$.    The 
Kronecker product between matrices $X$ and $Y$ is denoted by $X \otimes Y$.
If two vectors, $x_1$ and $x_2\in\R^n$ are orthogonal this is denoted by
$x_1 \perp x_2$.

\section{Data-Driven Control}
\label{s:ddcontrol}

\subsection{Problem Formulation}
\label{ss:formulation}

The problem to be considered is the control of a linear time-invariant system, $G$,
with input, $u(k) \in \R^{\nin}$, and output $y(k)\in\R^{\ny}$.  We assume that system
$G$ has finite order, $\nx$.   Rather than specify $G$ in terms of
a order-$\nx$ minimal state-space representation, our characterisation of $G$ is in
terms of a past record of input-output data over $K$ consecutive time-steps,
($\ud(k),\yd(k)$), $k = 1,\dots,K$.   
The superscript, $d$, is used to contrast the past data with current or future values of $u$ and $y$.
In the SMM development we assume that $\ud$ and $\yd$ are noise-free.

The predictive controller uses input and measured output signals from 
an immediate past horizon of length-$\Tp$ to calculate the optimal control 
input over a length-$\Tf$ horizon into the future.  The measured output
contains i.i.d.\ noise.  At every time-step $k$,
\begin{equation}
\label{e:ypmeas}
\ymeas(k)  =   y(k)  +  v(k),
\end{equation}
with $\Exp{v(k)} = 0$, and $\var{v(k)} = \Sigma_v$.
The length-$\Tp$ sequence of past inputs and outputs are written
as vectors,
\begin{equation}
\label{e:pastdefn}
\up(k) =  \begin{bmatrix} u(k-\Tp+1) \\ \vdots \\ u(k) \end{bmatrix}
\text{ and }
\yp(k)  =  \begin{bmatrix} y(k-\Tp+1) \\ \vdots \\ y(k) \end{bmatrix}.
\end{equation}
Future inputs and outputs are indexed from the next time step,
\begin{equation}
\label{e:futuredefn}
\uf(k)  =   \begin{bmatrix} u(k+1) \\ \vdots \\ u(k+\Tf) \end{bmatrix}
\quad\text{and}\quad 
\yf(k)  =   \begin{bmatrix} y(k+1) \\ \vdots \\ y(k+\Tf) \end{bmatrix}.
\end{equation}

The control problem is posed in terms of a cost function over the length-$\Tf$
future  horizon.  For simplicity we will consider a quadratic cost, defined by
$P \in \Splus^{\,\ny \Tf}$ and $R \in \Splus^{\,\nin \Tf}$, but more
general cost functions are easily handled.  

\begin{problem}[Data-driven predictive control]
\label{p:generic}
Given an immediate past input and measured output signals, $\up(k)$ and $\ypmeas(k)$,  
find a future input sequence, $\uf(k)$, solving
\begin{align}
\minimise_{\uf, \yf}  \quad & \| \yf \|_{\ycost}  +  \| \uf \|_{\ucost},\nonumber \\
\noalign{\medskip}
\st \quad & \begin{bmatrix} \yp \\ \yf \end{bmatrix}  =  G\left( \begin{bmatrix} \up \\ \uf \end{bmatrix} \right),
\quad \uf \in \Uset, \quad \yf \in \Yset \label{e:genericdyn}.
\end{align}
\end{problem}

In the standard receding horizon predictive control structure only the first time step of $\uf$
is applied to the plant.   At the next time-step new measurement data is available and the
entire problem is resolved.

The sets $\Uset$ and $\Yset$ denote input and output constraint sets, which are typically
assumed to be convex.    There are several difficulties in the above formulation.  To begin,
we have only the measured past output, $\ypmeas$,  but the characterisation of the
dynamics applies to (at best) the noise-free past output, $\yp$.   In addition, the
characterisation of the plant input-output mapping above is in a generic operator form.
Standard model predictive control assumes that
the future output, $\yf$, is characterised in terms of a measured state and the future
input, $\uf$.   Data-driven predictive control, on the other hand, uses the characterisation
of $\yf$ directly in terms of $\up$, $\yp$, and $\uf$.  The basis of this is the Willems' Lemma
described in the next section.

\subsection{Willems' Fundamental Lemma}
\label{ss:willems}

Our model of the system is specified in terms of a single, length-$K$, trajectory.
The extension to multiple trajectories is straightforward and given
in \cite{waarde:2020p}.   The input-output model will characterise all length-$T$ input-output
data sequences that can be generated by the system.

Hankel matrices, of dimension $(\nin T) \times M$  and dimension $(\ny T) \times M$ are created,
\[
\Hu = \begin{bmatrix} \ud(1) & \ud(2) & \cdots &  \ud(M)\\
				  \ud(2) & \ud(2) & \cdots & \ud(M+1) \\
				  \vdots & \vdots &      & \vdots \\
				  \ud(T) & \ud(T+1) & \cdots & \ud(M+T-1) 
	  \end{bmatrix}
\]
and
\[
\Hy = \begin{bmatrix} \yd(1) & \yd(2) & \cdots & \yd(M) \\
				  \yd(2) & \yd(3) & \cdots & \yd(M+1) \\
				  \vdots & \vdots &      & \vdots \\
				  \yd(T) & \yd(T+1) & \cdots & \yd(M+T-1) 
	  \end{bmatrix}.
\]
We assume that $T$ has been chosen so that $\ny\Tp \geq \nx$.
Furthermore $M$ has been chosen such that $M  \geq 2T(\nin + \ny)$.
The input sequence $u$ is assumed to be
persistently exciting of order at least $\nin T+\nx$.   These conditions
require that the data length satisfies,
$
K \geq M  +  T  -  1.
$

Under the persistency of excitation assumption
$\Hu$ has full row rank, equal here to $\nin T$.
Generically $\Hy$ also has full row rank (equal to
$\ny T$).

\begin{lemma}[Willems' Fundamental Lemma]
\label{l:willems}
Under the matrix-size and persistency
of excitation assumptions, the length-$T$ input-output pair ($u$,$y$) is
a trajectory of the system $G$, iff there exists $g\in\R^M$
such that,
\begin{equation}
\label{e:willems}
\begin{bmatrix}  u \\ y \end{bmatrix}
	 =  
	\begin{bmatrix} \Hu \\ \Hy \end{bmatrix}  g.
\end{equation}
\end{lemma}
See Theorem~1 in \cite{willems:2005c} and Lemma~2 in \cite{de-persis:2020a}.

It follows that the matrix of stacked Hankel matrices, $\begin{bmatrix} \Hu^T & \Hy^T \end{bmatrix}^T$,
has rank equal to $\nin T + \nx$.  The if and only if nature of the result
means that the matrix multiplication in~(\ref{e:willems}) can be used
as a complete characterisation of the trajectories specified by Equation~\ref{e:genericdyn}
in Problem~\ref{p:generic}.

\subsection{Data-Driven Predictive Control: DeePC}
\label{ss:deepc}

To apply~(\ref{e:willems}) to predictive control we partition the signal sequences into
past and future, as given in~(\ref{e:pastdefn}) and~(\ref{e:futuredefn}), satisfying
$T  = \Tp + \Tf$.
We also partition the $\Hu$ and $\Hy$ correspondingly,
\[
\renewcommand{\arraystretch}{1.25}
\Hu = \left[  \begin{array}{c}   \Hup \\
					  \hline 
					    \Huf 
					    \end{array} \right],
\text{ with } \Hup \in \R^{(\nin\Tp) \times M}, \Huf \in \R^{(\nin\Tf) \times M},
\]
and
\[
\renewcommand{\arraystretch}{1.25}
\Hy = \left[ \begin{array}{c}   \Hyp \\
					  \hline 
					    \Hyf 
					    \end{array} \right],
\text{ with } \Hyp \in \R^{(\ny\Tp) \times M}, \Hyf \in \R^{(\ny\Tf) \times M}.
\]
We assume that this partition satisfies $\ny \Tp \geq \nx$ so that
the assumptions in Lemma~\ref{l:willems} also apply to $\Hyup$ (defined subsequently).

As reordering the rows of both $u$ and $y$ commensurately with reordering the rows of the
stacked Hankel matrices in~(\ref{e:willems}) does not affect $g$, we can replace~(\ref{e:willems})
with the following characterisation of the dynamics.
\begin{equation}
\label{e:willems1}
\begin{bmatrix}  \up \\ \yp \\ \uf \\ \yf \end{bmatrix}
 = 
	\begin{bmatrix} \Hup \\ \Hyp \\ \Huf \\ \Hyf \end{bmatrix}
	g
\end{equation}	

Applying this characterisation of the dynamics in Problem~\ref{p:generic} gives the DeePC problem formulation.

\begin{problem}[DeePC]
\label{p:deepc}
Given an immediate past input and measured output signals, $\up(k)$ and $\ypmeas(k)$,  
find a future input sequence, $\uf(k)$, solving
\begin{align}
\minimise_{g, \uf, \yf, \sigma_y}  \quad & \| \yf \|_{\ycost}  +  \| \uf \|_{\ucost}  +  \lambda_1 \|\sigma_y\|^2_2
		 +  \lambda_2 \rho(g) \nonumber \\
\noalign{\medskip}
\st \quad & \begin{bmatrix}  \up \\ \ypmeas - \sigma_y \\ \uf \\ \yf \end{bmatrix}
			 = 
			\begin{bmatrix} \Hup \\ \Hyp \\ \Huf \\ \Hyf \end{bmatrix} g, \label{e:deepcdyn}\\
	&  \uf \in \Uset, \quad \yf \in \Yset. \nonumber
\end{align}
\end{problem}

The DeePC formulation above has several variants all of which differ in significant ways from the
generic Problem~\ref{p:generic}.   One difference is that as $\ypmeas$ contains noise,  it is
almost certain that,
\[
\begin{bmatrix} \up \\ \ypmeas \end{bmatrix}  \notin  
		 \range \left( \begin{bmatrix} \Hup \\ \Hyp \end{bmatrix} \right).
\]
The variable $\sigma_y$ acts as a surrogate for this noise, making~(\ref{e:deepcdyn}) solvable, and its
size minimisation is included as a regularisation term, $\lambda_1\|\sigma_y\|_2^2$.

The other aspect is that $g$ plays a dual role.  The first two block rows of~(\ref{e:deepcdyn}) are constraining
$g$ such that the initial conditions (specified by the immediate past data) are close to being satisfied.   The
second role is the use of $g$ in finding the optimal $\uf$ and $\yf$.   The potential conflict can be seen by noting
that $g$ may select a ``best-case'' noise $\sigma_y$ from the point of view of the minimisation of 
the quadratic cost.   The regularisation of $g$, specified by $\lambda_2 \rho(g)$, is used
to trade-off between these potential conflicting roles of $g$.   In the case where $\Hy$ is actually 
created from noisy measurements the matrix of stacked past Hankel matrices is full rank and~(\ref{e:deepcdyn})
does not impose any constraint on the initial condition matching (see \cite{mattsson:2023a}).  In this case the 
regularisation is essential to achieving good performance with DeePC.

The more recent data-driven predictive control formulations, $\gamma$-DDPC and GDPC, improve upon the
DeePC formulation by compressing the representation and partially separating the initial condition matching and
predictive control parameterisations.   Regularisations are still required although \cite{verheijen:2023a} provides
simulations that illustrate that they can outperform DeePC.

\section{Optimal Multi-step Prediction}

Our formulation is conceptually based on the idea that we will separate the $g$ vector in~(\ref{e:willems1})
into two orthogonal parts,
\begin{equation}
\label{e:gpgf}
\begin{bmatrix}  \up \\ \yp \\ \uf \\ \yf \end{bmatrix}
 = 
	\begin{bmatrix} \Hup \\ \Hyp \\ \Huf \\ \Hyf \end{bmatrix}
	(\gp  +  \gf),
\end{equation}
where $\gp$ matches the initial conditions specified by $\up$ and $\yp$, and
$\gf$ specifies the mapping between $\uf$ and $\yf$.  While $\gp$ also specifies a future
input-output trajectory, $\Hup\gf = \Hyp\gf = 0$.  Furthermore $\gp \perp \gf$ and so $\gf$
can completely determine the $\uf$, $\yf$ mapping for the predictive control step.   
See~\cite{dorfler:2023a} for the use of a similar concept for DeePC regularisation.

\subsection{Derivation of a Parsimonious Formulation}
\label{ss:derivation}

The mapping in~(\ref{e:willems1}) will be condensed into a minimal parameterisation,
similar in motivation to the parameterisations used in $\gamma$-DDPC and GDPC.

Consider stacked Hankel matrices representing the past
and future input-output sequences,
\[
\Hyup =  \begin{bmatrix} \Hup \\ \Hyp \end{bmatrix},
\text{ and }
\Hyuf  =  \begin{bmatrix} \Huf \\ \Hyf \end{bmatrix}.
\]
The matrix of immediate past data, $\Hyup$, is factorised 
into,
\[
\Hyup  =  \Lp  \Qp^T, 
\]
where $\Lp \in \R^{((\nin+\ny)\Tp) \times M}$ is lower triangular
and $\Qp \in \R^{M\times M}$ is unitary.
This can be further partitioned into,
\begin{equation}
\label{e:HyupLQ}
\renewcommand{\arraystretch}{1.25}
\begin{bmatrix} \Hup \\ \Hyp \end{bmatrix}
= 
	\left[ \begin{array}{ c | c | c}
	 \Lup  &  0   &  0 \\
	 \hline
	 \Lyup & \Lyp & 0 \\
	\end{array} \right]
	 \left[  \begin{array} {ccc}
	  & \Qup^T &  \\
	 \hline
	   & \Qyp^T &  \\
	 \hline
	   & \Qnp^T &  
	 \end{array}\right],
\end{equation}
where both $\Lup \in \R^{(\nin \Tp) \times (\nin \Tp)}$ and $\Lyp \in \R^{(\ny \Tp) \times \nx}$
are lower triangular.  The fact that $\Lyp$ has only $\nx$ columns follows from the observation
that Lemma~\ref{l:willems} specifies that $\rank(\Hyup) = \nin \Tp + \nx$ and the persistency of
excitation gives $\rank(\Lup) = \nin \Tp$.   This also implies that the dimension of the null space of $\Hyup$
(which is also the column dimension of $\Qnp$) is $M - \nin \Tp - \nx$.  

We can use this factorisation to simplify our representation of all length-$\Tp$ sequences that could
have been generated by the system.  By solving for $\xu \in \R^{\nin \Tp}$  and $\xy \in \R^{\nx}$
satisfying,
$
\up  =  \Lup \xu
$
and
$
\yp  = \Lyup \xu  +  \Lyp \xy,
$
we can rewrite the past data part of Willems' condition as,
\[
\begin{bmatrix} \up \\ \yp \end{bmatrix}
	 =  
	\left[ \begin{array}{ c | c }
		 \Lup  &  0   \\
		 \hline
	 	\Lyup & \Lyp  \\
		\end{array} \right] 
	\begin{bmatrix} \xu \\ \xy \end{bmatrix}.
\]	
While we will parametrise the solutions in terms of $\xu$ and $\xy$, 
it also possible to write $\gp$ in~(\ref{e:gpgf}) as,
\[
\gp  =  \begin{bmatrix} \Qup & \Qyp \end{bmatrix}
			\begin{bmatrix} \xu \\ \xy \end{bmatrix}.
\]
Note that all $g$ in the null-space of $\Hyup$ are given
by $g = \Qnp \xnp$ where $\xnp \in \R^{M - \nin \Tp - \nx}$
and that $\gp$ is orthogonal to every column of $\Qnp$.   We will
use this orthogonal decomposition to characterise all possible future
input-output trajectories that match a specific past input-output trajectory.
The specific $\gp$ for a given $\up$, $\yp$ trajectory depends of course
on the trajectory.  However the null-space parameterisation $\Qnp \xnp$ 
is independent of the past trajectory being matched.   The consequence of
this independence is that the past input-output trajectory affects the future $\yf$
(as we would expect from the dynamics) and also the future $\uf$.

We will now apply the LQ decomposition approach to the Hankel matrices
that generate the length-$\Tf$ future input-output trajectories.   As we want
these to also match a past length-$\Tp$ input-output trajectory we will constrain
the choice of $\gf$ to those that lie within the null-space of $\Hyup$.  

To do this the LQ factorisation is applied to $\Hyuf \Qnp$ and
\begin{equation}
\label{e:LfQf}
 \Hyuf \Qnp   =   \begin{bmatrix} \Huf \\ \Hyf \end{bmatrix}\Qnp
	 =  \Lf \Qf^T,
\end{equation}
gives a lower triangular $\Lf \in \R^{((\nin + \ny)\Tf) \times (M - \nin \Tp - \nx)}$ and a unitary
$\Qf \in \R^{(M - \nin \Tp - \nx) \times (M - \nin \Tp - \nx)}$.  If $\Qnp$  has at least
$(\nin + \ny)\Tf$ columns,  Equation~\ref{e:LfQf} can be partitioned conformally 
with the future inputs and outputs,
\begin{equation}
\label{e:Lufdefn}
\Hyuf \Qnp
 =  \left[ \begin{array}{c | c}
   \Luf &  0 \\
   \hline
   \Lyuf & \Lyf
   \end{array} \right]
   		\left[
		\renewcommand{\arraystretch}{1.25}
		\begin{array} {c}
			\Qyuf^T \\
		\hline
			\Qnf^T 
		\end{array} \right],
\end{equation}
where $\Luf \in \R^{(\nin \Tf) \times (\nin \Tf)}$ and $\Lyuf \in \R^{(\ny \Tf) \times (\nin \Tf)}$.
Note that the $\Qnp$ column dimension assumption is satisfied for any $\nx \leq \ny\Tp$
if $M \geq 2T(\nin + \ny)$.
The following lemma, proven in the Appendix, illustrates several interesting structural properties of the decomposition
in~(\ref{e:Lufdefn}).

\begin{lemma}
\label{l:LfQf}
The decomposition in~(\ref{e:Lufdefn}) satisfies:\\
\hspace*{\parindent}{\it a)} $\rank(\Luf) = \nin \Tf$  and {\it b)}  $\Lyf = 0$.
\end{lemma}

We can now reduce the dimension of the $\Hyuf\Qnp$ parametrisation to the column
dimension of $\Qyuf$.  Consider
$
\gf  = \Qnp \Qyuf  z,
$
with
$z \in \R^{\nin \Tf}$,
which results in,
\[
\Hyuf \, \gf  =  \begin{bmatrix} \Luf \\ \Lyuf \end{bmatrix} z.
\]
Because $\Luf$ is full rank, any specified future input will uniquely determine $z$.  This in turn
specifies a unique future output via $\Lyuf$.  However we must also account for the contribution of
 $\xu$ and $\xy$ to the future input and output.

We can now reformulate~(\ref{e:willems1}) more parsimoniously as
\begin{equation}
\label{e:LQpf_defn}
\begin{bmatrix}  \up \\ \yp \\ \uf \\ \yf \end{bmatrix}
 = 
	\begin{bmatrix} \Lup   & 0        & 0  \\
	                         \Lyup & \Lyp & 0 \\
	                         \Suu & \Suy & \Luf \\
	                         \Syu & \Syy & \Lyuf \end{bmatrix}
	 \begin{bmatrix}  \xu \\ \xy \\ z \end{bmatrix},
\end{equation}	
where we have defined the $S$ matrices via
\[
\begin{bmatrix}  \Suu & \Suy \\
			\Syu & \Syy \end{bmatrix}
	 =  \begin{bmatrix} \Huf \\ \Hyf \end{bmatrix}
		\begin{bmatrix} \Qup & \Qyp \end{bmatrix}.
\]
Note that the lower triangular matrix in~(\ref{e:LQpf_defn}) is full column rank, 
with rank equal to $\nin T + \nx$. 

\subsection{Optimal multi-step predictor}
\label{ss:opt_pred}

We will now apply the parametrisation in~(\ref{e:LQpf_defn}) to a $\Tf$-step ahead
prediction of the future output.  Optimality in this context refers to finding the minimum
variance unbiased predictor of the future output.
To begin we pose the prediction problem.

\begin{problem}
\label{p:Tfstep_simple}
Given a length-$\Tp$ input-output signal sequence, $\up$, $\ypmeas$, generated by 
the system $G$ with output measurement noise specified by~(\ref{e:ypmeas}),
and given a specified contiguous length-$\Tf$ future input
sequence, $\uf$, estimate the future system output, $\yf$ (with the estimate
denoted by $\yfest$).
\end{problem}

Consider the following linear form of the estimator,
\begin{equation}
\label{e:predictor}
\yfest = \Eup \up  +  \Eyp \ypmeas  +  \Euf \uf,
\end{equation}
where the predictor matrices are given by,
\begin{align*}
\Euf  & =   \Lyuf \Luf^{-1}, \quad \Eyup = \Lyup \Lup^{-1} , \quad \Psixy = \Syy - \Euf^{-1}\Suy,\\
\Eup  & =   (\Syu - \Euf \Suu) \Lup^{-1} - \Psi \Exy \Eyup,\\
\Eyp & =   \Psi \Exy, 
\end{align*}
and the matrix $\Exy$ is given by,
\[
\Exy  =  \left( \Lyp^T \SigmaV^{-1} \Lyp \right)^{-1}
				\Lyp^T  \SigmaV^{-1},
\]
where  $\SigmaV = \ITp \otimes \Sigma_v$.  
Note that the predictor matrices $\Eup$, $\Eyp$, and $\Euf$, can be calculated
offline.

In SPC a linear predictor of the form in~(\ref{e:predictor}) is obtained as the
least-squares solution to the data matrix equation,
\[
\Hyf =  \Eup \Hup +  \Eyp \Hyp  + \Euf \Huf.
\]
This is a good heuristic but not optimal.  In contrast...

\begin{theorem}
\label{t:blueyf}
Given:  an LTI system, $G$, uniquely specified by SMM matrices; 
a length-$\Tp$ immediate past data sequence ($\up$,$\yp$);  and
a future input, $\uf$.   The estimate $\yfest$, given in~(\ref{e:predictor}),
is the best linear unbiased estimate (BLUE) of $\yf$.  
Furthermore the covariance of $\yfest$ is equal to
\[
\cov(\yfest) = \Psi (\Lyp^T\SigmaV^{-1} \Lyp)^{-1} \Psi^T.
\]
\end{theorem}

We now have an optimal (BLUE) estimator to apply to the MPC problem.


\section{Signal Matrix Model Predictive Control}
\label{s:smmpc}

One can immediately apply this approach to an MPC 
problem with~(\ref{e:predictor}) specifying the dynamics.  The following MPC problem
is an alternative to the DeePC approach given in Problem~\ref{p:deepc}.

\begin{problem}[SMMPC]
\label{p:smmpc}
Given an immediate past input and measured output signals, $\up$ and $\ypmeas$,  
find a future input sequence, $\uf$, solving

\begin{align*}
\minimise_{\uf, \yfest}  \quad & \| \yfest \|_{\ycost}  + \| \uf \|_{\ucost}  \\
\noalign{\medskip}
\st \quad &  \yfest  =  \Eup \up  +  \Eyp \ypmeas  +  \Euf \uf,\\
	&  \uf \in \Uset, \quad \yfest \in \Yset.
\end{align*}
\end{problem}

In contrast to DeePC, $\gamma$-DDPC, and GDPC, this formulation does not 
require the designer to select regularisation parameters.   It also has fewer 
optimisation variables and as $\Eup$, $\Eyp$ and $\Euf$ are calculated offline
it will run more quickly.   As with almost all MPC formulations it is common practice
that the constraint $\yfest \in \Yset$ is specified with a penalised slack variable
to avoid potential infeasibility.

\subsection{Simulation demonstration}

The flight control benchmark problem in~\cite{verheijen:2023a} is used to demonstrate
the approach\footnote{{\sc Matlab} code
running this comparison is publicly available at
 {\tt https://doi.org/10.3929/ethz-b-000663423}}.
The model data experiment, $\{\ud,\yd\}$, is of length $K=2500$ and, in contrast to the theoretical
development in prior sections, contains noise with a variance of $\Sigma_v = 0.25^2\Inu$. 
To create the SMM model structure
in~(\ref{e:LQpf_defn}) we take the maximal rank of the past data as the state dimension, $\nx = \ny\Tp$.
The horizon lengths are $\Tp = \Tf = 40.$   The same noise variance is used for the measurement
noise on $\yp$ in the step response simulations.  For a comparison we use the MPC plus Kalman
filter design in~\cite{verheijen:2023a}.  The subspace identification function {\tt n4sid}, truncated to
the true system order, is used to give the state-space model. 

\begin{figure}
\includegraphics[width=\columnwidth]{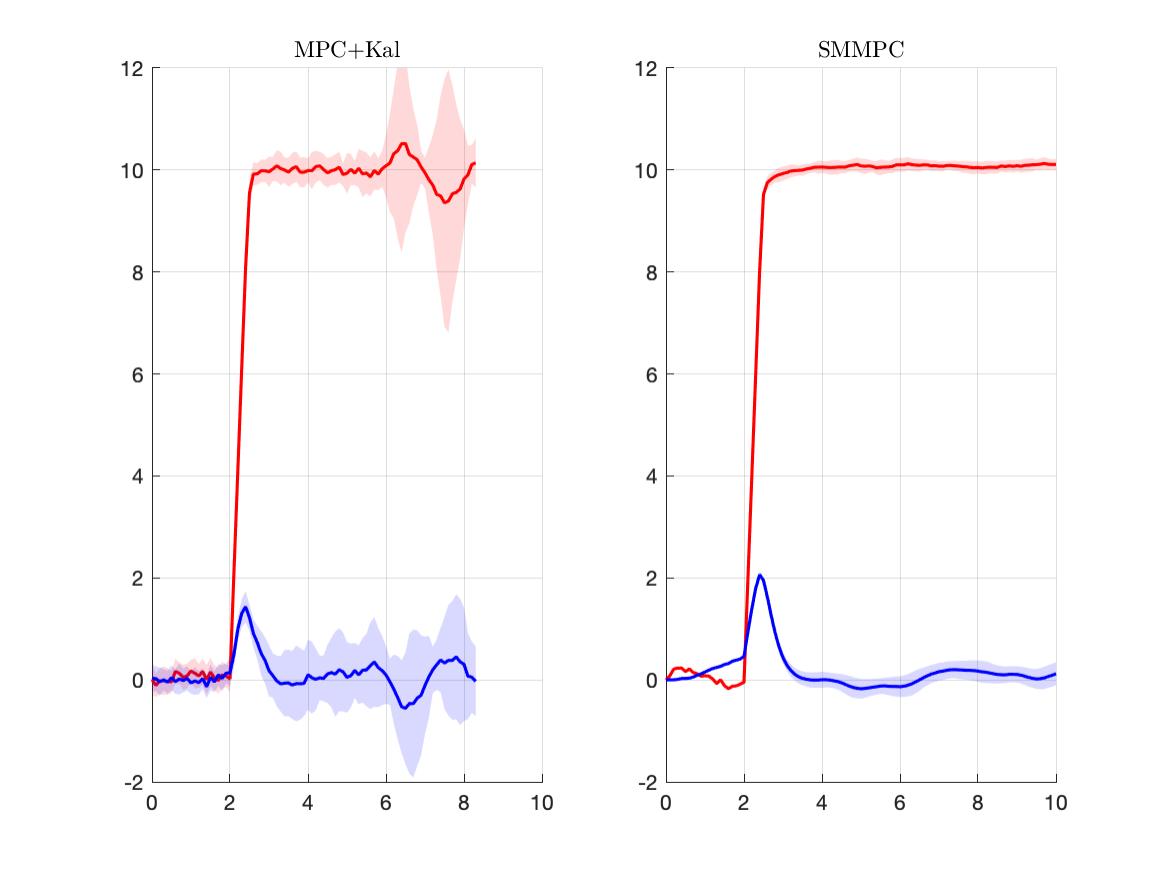}
\caption{\label{f:outputs} Output trajectories, mean value and 1 std.\ dev.\ (shaded)}
\end{figure}

\begin{figure}
\includegraphics[width=\columnwidth]{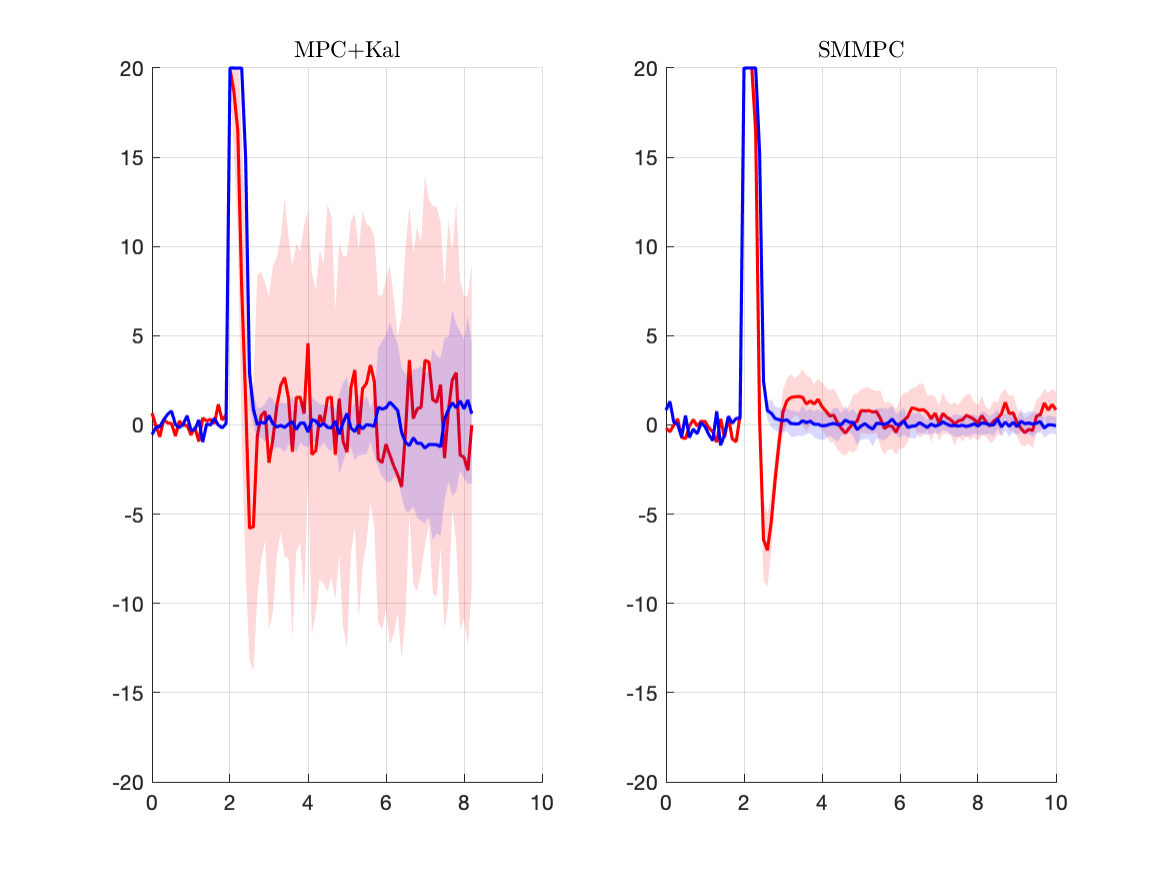}
\caption{\label{f:inputs} Input trajectories, mean value and 1 std.\ dev.\ (shaded)}
\end{figure}

Figure~\ref{f:outputs} illustrates the mean and standard deviation of a step response over
30 Monte Carlo runs (rebuilding the noisy models with each run).  The corresponding input signals
are given in Figure~\ref{f:inputs}.  

\begin{figure}
\includegraphics[width=\columnwidth]{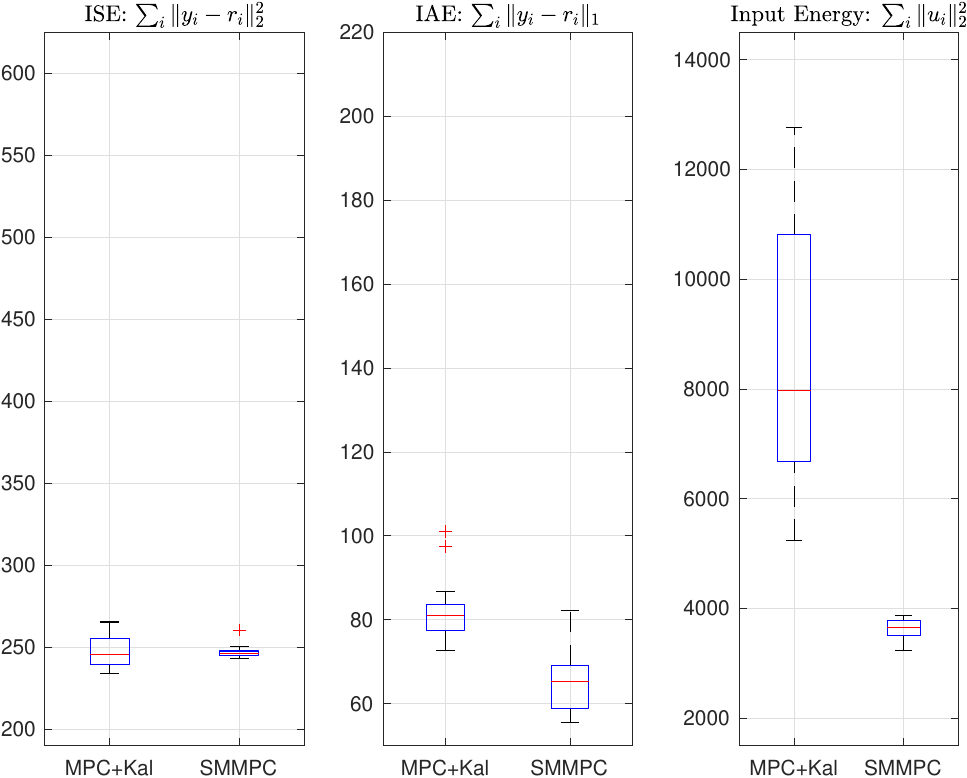}
\caption{\label{f:perf} Performance indices over 30 Monte Carlo runs}
\end{figure}

Three performance indices are shown in Figure~\ref{f:perf}.  The axes are chosen as
close as possible to those in~\cite{verheijen:2023a} to facilitate comparison.
The SMMPC method performs as well or better than the other methods.  

The MPC optimisation problems were solved with OSQP~\cite{stellato:2020a}.  The
SMMPC calculation time is similar to the Kalman filter plus MPC computation time.


\section{Conclusions}

As is clear from Willems' fundamental lemma the SMM matrices specify a
unique causal LTI system of finite order.  The model specification in terms of
past input-output trajectories avoids the need to specify a state representation.
But of course the SMM matrix representation is equivalent, in terms of input-output
responses, to a state-space representation.  So the strong similarities, such as
the dimension of $\xy$ being equal to the system order, should not be 
surprising.  

Another defining characteristic of state-space systems is that the entire past
input-output behaviour affects the future input-output behaviour only via the
initial state.
A similar thing occurs here as $\up$ and $\yp$ completely determine $\xu$ and
$\xy$ which in turn uniquely specify the future input-output relationship.
The difference in the SMM case is that we use $\xu$ and $\xy$ to specify a 
future input as well as a future output.  This contrasts with state-space representations
where the initial state specifies the future output corresponding to a future 
input of zero.   The benefit of the SMM approach here is that it allows us to
maintain linearity in the input-output relationships making it simpler to 
formulate an optimal prediction problem.   However the part of the future
input arising from $\xu$ must be taken into consideration when selecting
the future input for predictive control purposes.  This is done via~(\ref{e:predictor}).

The optimal prediction problem given here is easily solved because the 
underlying system is causal and the dynamics are linear.   Regularisation
is not required to trade-off between the prediction and control parts of the
problem as they are decoupled.   However this doesn't necessarily mean that
regularisation is not beneficial.    The recent developments in using
kernel-based regularisers for identification have given significant benefits 
in the system identification domain~\cite{pillonetto:2010d} and could be
applied here to reduce noise in $\Hy$ when the signals come from experimental
data. 

Note also that the variance of the predictor (Theorem~\ref{t:blueyf})
depends directly on the experimental data used to generate the SMM
matrices $\Hu$ and $\Hy$.   This makes it clear that the issues arising in
experiment and input design for system identification also arise here.

\appendix
\subsection{Proof of Lemma~\ref{l:LfQf}:}

To show {\it a)} note that the persistency of excitation assumption
implies that $\Hu$ has rank $\nin T$ and so $\rank(\Luf) = \rank(\Huf)$ 
as $\nin T - \nin \Tp = \nin \Tf$.

To show {\it b)} Assume that $\Lyf \neq 0$ and choose a vector 
$\hat{w} \in \R^{M - \nin T - \nx}$ such that $\Lyf \hat{w} \neq 0$.
Calculate the vector $\hat{g} \in \R^{M}$ via
$
\hat{g}  =  \Qnp \Qnf \hat{w}.
$
Now consider the input-sequence corresponding to,
\[
\begin{bmatrix} \Hyup \\ \Hyuf \end{bmatrix} \hat{g}
  =   \begin{bmatrix} 0 \\ \Hyuf \hat{g}\end{bmatrix} 
  =  \begin{bmatrix} 0 \\ \Lf \Qf^T \Qnf \hat{w}\end{bmatrix} 
 =  \begin{bmatrix} 0 \\  \begin{bmatrix} 0 \\ \Lyf \hat{w} \end{bmatrix}\end{bmatrix}.
\]
This corresponds to an input-output sequence with 
$\up = \uf = \yp = 0$, $\yf = \Lyf\hat{w}$.  
This contradicts the uniqueness assumption on the system $G$ as $\yf = 0$ is also a solution 
to this case.  Therefore $\Lyf = 0$.
\qed

\subsection{Proof of Theorem~\ref{t:blueyf}}
To remove the affine deterministic terms define 
\[
\zetaf = \yf -  \Eup \up - \Euf \uf,
\text{ and }
\nup = \yp - \Eyup \up,
\]
and note that the noise-free signals satisfy,
\[
\zetaf = \Psixy \xy \text{ and } \nup = \Lyp \xy.
\]
The BLUE of $\xy$ is,
$
\xyest = \Exy \nupmeas,
$
where $\nupmeas = \ypmeas - \Eyup \up$.
The estimator in~(\ref{e:predictor}) can be reformulated as
\begin{equation}
\label{e:zetaBLUE}
\zetafest = \Psi \xyest,
\end{equation}
and as it is a linear scaling of BLUE of $\xy$, it is the BLUE
of $\zetafest$.   The covariance is given by
\[
\cov{\yfest} = \cov{\zetafest} = \Psi (\Lyp^T \SigmaV^{-1} \Lyp)^{-1} \Psi^T.
\]

To show that this is the BLUE consider all linear estimators, 
$
\zetafest = Z^T \nupmeas.
$
To be unbiased they must satisfy
$
\Exp{Z^T \nupmeas} = \zetaf,
$
or
\[
Z^T  \Exp{\nupmeas} - \Psi \xy = 0, \quad\text{for all }\xy.
\]
As $\Exp{\nupmeas} = \Lyp \xy$, we must have,
$
Z^T\Lyp - \Psi = 0.
$
This implies that $\range(Z^T) = \range(\Psi)$ and $Z^T$ can be
expressed as a factorisation,
$Z^T = \Psi F$ with $F \Lyp = \Inx$.

The covariance of the $Z^T$ estimator is given by $\cov(\zetafest) = Z^T \SigmaV Z$.
The difference with respect to the estimator in~(\ref{e:zetaBLUE}) is
\begin{align*}
Z^T \SigmaV Z & - \Psi (\Lyp^T \SigmaV^{-1} \Lyp)^{-1} \Psi^T \\
	& = \Psi F \SigmaV F^T \Psi^T - \Psi (\Lyp^T \SigmaV^{-1} \Lyp)^{-1} \Psi^T,\\
\noalign{\hspace{-12pt}and by exploiting the fact that  $\Inx = F\Lyup,$}
	& = \Psi F (\SigmaV - \Lyp (\Lyp^T \SigmaV^{-1} \Lyp)^{-1} \Lyup^T ) F^T \Psi^T \\
	& = \Psi F \Theta \SigmaV^{-1} \Theta^T  F^T \Psi^T  \geq 0,
\end{align*}
where $\Theta  = \Sigma_V - \Lyp(\Lyp^T \SigmaV^{-1} \Lyp )^{-1} \Lyp^T$.  Hence the
estimator in~(\ref{e:zetaBLUE}) achieves the minimum  covariance.	
\qed




\begin{thebibliography}{10}
\providecommand{\url}[1]{#1}
\csname url@samestyle\endcsname
\providecommand{\newblock}{\relax}
\providecommand{\bibinfo}[2]{#2}
\providecommand{\BIBentrySTDinterwordspacing}{\spaceskip=0pt\relax}
\providecommand{\BIBentryALTinterwordstretchfactor}{4}
\providecommand{\BIBentryALTinterwordspacing}{\spaceskip=\fontdimen2\font plus
\BIBentryALTinterwordstretchfactor\fontdimen3\font minus
  \fontdimen4\font\relax}
\providecommand{\BIBforeignlanguage}[2]{{%
\expandafter\ifx\csname l@#1\endcsname\relax
\typeout{** WARNING: IEEEtran.bst: No hyphenation pattern has been}%
\typeout{** loaded for the language `#1'. Using the pattern for}%
\typeout{** the default language instead.}%
\else
\language=\csname l@#1\endcsname
\fi
#2}}
\providecommand{\BIBdecl}{\relax}
\BIBdecl

\bibitem{verheijen:2023a}
P.~Verheijen, V.~Breschi, and M.~Lazar, ``Handbook of linear data-driven
  predictive control: {Theory,} implementation and design,'' \emph{Annual
  Reviews in Control}, vol.~56, p. 100914, 2023.

\bibitem{berberich:2021a}
J.~Berberich, J.~K\"{o}hler, M.~A. M\"{u}ller, and F.~Allg\"{o}wer,
  ``Data-driven model predictive control with stability and robustness
  guarantees,'' \emph{IEEE Trans.\ Automatic Control}, vol.~66, no.~4, pp.
  1702--1717, Apr. 2021.

\bibitem{favoreel:1999a}
W.~Favoreel, B.~De~Moor, and M.~Gevers, ``{SPC: Subspace} predictive control,''
  in \emph{Proc.\ IFAC World Congress}, 1999, pp. 4004--4009.

\bibitem{willems:2005c}
J.~C. Willems, P.~Rapisarda, I.~Markovsky, and B.~L. De~Moor, ``A note on
  persistency of excitation,'' \emph{Systems \& Control Letters}, vol.~54,
  no.~4, pp. 325--329, 2005.

\bibitem{fiedler:2021a}
F.~Fiedler and S.~Lucia, ``On the relationship between data-enabled predictive
  control and subspace predictive control,'' in \emph{Proc.\ European Control
  Conference}, 2021, pp. 222--229.

\bibitem{coulson:2018a}
J.~Coulson, J.~Lygeros, and F.~D\"{o}rfler, ``Data-enabled predictive control:
  In the shallows of the {DeePC},'' in \emph{Proc.\ European Control
  Conference}, November 2019, pp. 307--312.

\bibitem{breschi:2023a}
V.~Breschi, M.~Fabris, S.~Formentin, and A.~Chiuso, ``Uncertainty-aware
  data-driven predictive control in a stochastic setting,'' in \emph{IFAC
  PapersOnLine}, vol. 56-2, 2023, pp. 10\,083--10\,088.

\bibitem{breschi:2023b}
V.~Breschi, A.~Chiuso, and S.~Formentin, ``Data-driven predictive control in a
  stochastic setting: a unified framework,'' \emph{Automatica}, vol. 152, p.
  110961, 2023.

\bibitem{lazar:2023a}
M.~Lazar and P.~Verheijen, ``Generalized data-driven predictive control:
  {Merging} subspace and {Hankel} predictors,'' \emph{Mathematics}, vol.~11, p.
  2216, 2023.

\bibitem{Yin:2023a}
M.~Yin, A.~Iannelli, and R.~S. Smith, ``Maximum likelihood estimation in
  data-driven modeling and control,'' \emph{IEEE Trans.\ Automatic Control},
  vol.~68, no.~1, pp. 317--328, 2023.

\bibitem{yin:2021a}
M.~Yin and R.~S. Smith, ``On low-rank {Hankel} matrix denoising,'' in
  \emph{IFAC-PapersOnLine (System Identification Symposium)}, vol.~54, no.~7,
  2021, pp. 198--203.

\bibitem{waarde:2020p}
H.~J. van Waarde, C.~De~Persis, M.~K. Camlibel, and P.~Tesi, ``Willems'
  fundamental lemma for state-space systems and its extension to multiple
  datasets,'' \emph{IEEE Control Systems Letters}, vol.~4, no.~3, pp. 602--607,
  2020.

\bibitem{de-persis:2020a}
C.~De~Persis and P.~Tesi, ``Formulas for data-driven control: Stabilization,
  optimality and robustness,'' \emph{IEEE Trans.\ Automatic Control}, vol.~65,
  no.~3, pp. 909--924, 2020.

\bibitem{mattsson:2023a}
P.~Mattsson and B.~Sch\"{o}n, Thomas, ``On the regularization in {DeePC},'' in
  \emph{Proc. IFAC World Congress, IFAC PapersOnLine}, vol. 56-2, 2023, pp.
  625--631.

\bibitem{dorfler:2023a}
F.~D\"{o}rfler, J.~Coulson, and I.~Markovsky, ``Bridging direct and indirect
  data-driven control formulations via regularizations and relaxations,''
  \emph{IEEE Trans.\ Automatic Control}, vol.~68, no.~2, pp. 883--897, 2023.

\bibitem{stellato:2020a}
B.~Stellato, G.~Banjac, P.~Goulart, A.~Bemporad, and S.~Boyd, ``{OSQP}: An
  operator splitting solver for quadratic programs,'' \emph{Mathematical
  Programming Computation}, vol.~12, no.~4, pp. 637--672, 2020.

\bibitem{pillonetto:2010d}
G.~Pillonetto and G.~De~Nicolao, ``A new kernel-based approach for linear
  system identification,'' \emph{Automatica}, vol.~46, pp. 81--93, 2010.

\end{thebibliography}
\end{document}